\documentclass[12pt]{article}

\usepackage{graphicx}
\usepackage{rangecite}

\def\be{\begin{equation}}
\def\ee{\end{equation}}
\def\bear{\begin{eqnarray}}
\def\eear{\end{eqnarray}}
\def\nn{\nonumber}

\def\half{{{1\over 2}}}

\def\a{{\alpha}}

\begin{document}


\begin{titlepage}
\vskip 1in
\begin{center}
{\Large
{Into the bulk: reconstructing spacetime from the c=1 matrix model}}
\vskip 0.5in
{Joanna L. Karczmarek and Daoyan Wang}
\vskip 0.3in
{\it 
Department of Physics and Astronomy\\
University of British Columbia
Vancouver, Canada}
\end{center}

\vskip 0.5in
\begin{abstract}
We write down exact solutions in the collective field theory of the c=1 matrix model and in dilaton-gravity coupled to a massless scalar.  Using the known correspondence between these two theories at the null boundaries of spacetime, we make a connection  between scalar fields in these two theories in the bulk of spacetime.  In the process, we gain insight into how a theory containing gravity can be equivalent to one without gravity.  We analyze a simple time-dependent background as an example.
\end{abstract}
\end{titlepage}





\section{Introduction}
\label{s1}

One of the hallmarks of gauge/gravity correspondence is the 
emergent nature of a non-compact spacetime dimension,
with the gauge theory dual living on the boundary of the gravitational
---or string theory--- spacetime.
The situation with c=1 string theory is somewhat similar.
On the string/gravity side, we have Liouville string theory,
which is a sub-critical string theory with a target spacetime
of dimension two.  The corresponding effective spacetime
action is dilaton-gravity coupled to a massless scalar
known as the `tachyon'.  On the gauge side, we have
a gauged Matrix Quantum Mechanics in a double scaling
limit, a large N system living in one dimension: time.
Using the collective field formalism, this MQM can be 
rewritten as a 1+1 dimensional field theory of a single
scalar, corresponding to the density of matrix eigenvalues.
It can be said that the spacial dimension in which Liouville 
strings propagate is emergent, as it is generated from the matrix 
eigenvalues.  It is less clear how gravity emerges in this picture.

\begin{figure}
\includegraphics[scale=0.9]{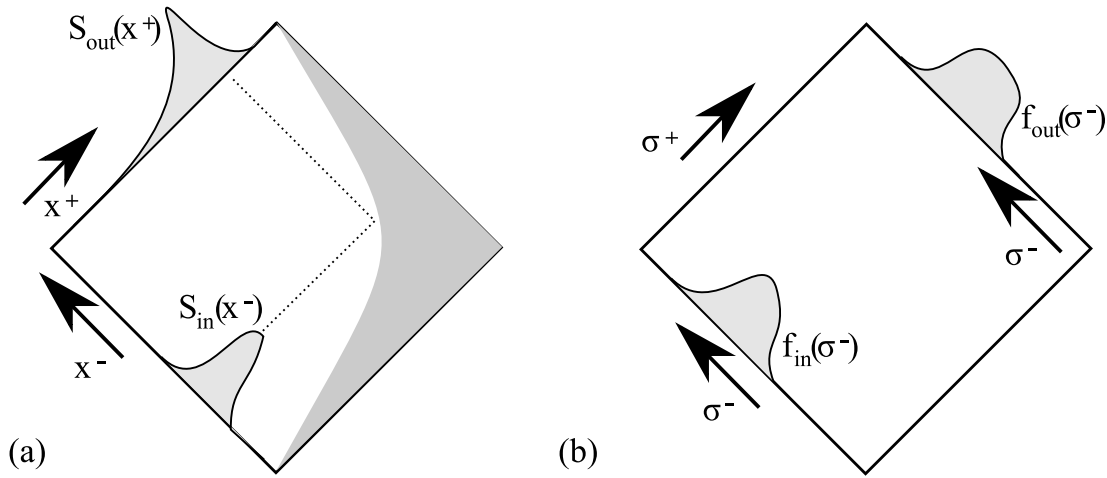}
\caption{{\bf (a)} Penrose diagram of Liouville string theory.
The shaded region is inaccessible to string degrees of freedom
due to the presence of the tachyon wall.  Massless tachyon
pulses scatter from $\cal{I}$$^-$ to $\cal{I}$$^+$.  The
shape of the outgoing pulse encodes information about
the interior metric and dilaton fields.
{\bf (b)} Penrose diagram of the collective field theory
for $\mu > 0$.}
\label{f1}
\end{figure}

The purpose of this paper is to explore in some detail how
gravity arises from the c=1 matrix model when the so called
`leg-pole' factors are taken into account, extending the 
results of \cite{Natsuume:1994sp}. We will think about
this construction as a toy model for holography, as follows.  
The spacetime of Liouville string theory (see Figure \ref{f1}(a))
is flat, and can be parametrized by two coordinates
$x$ and $t$, or $x^\pm = t \pm x$.  The string coupling varies with 
space like $g_{s} \sim \exp(2x)$, and the strong coupling region
at large $x$ is shielded by the presence of a tachyon background,
$T_0 \sim \exp(2x)$ which repels strings away from this region.
In addition, the same quantum improvement which leads to the 
inhomogeneous string coupling (and which is necessary in a 
noncritical string theory) also makes the tachyon massless.  
Finally, in two target
space dimensions, there are no transverse oscillators in the 
quantization of the string world-sheet, so the tachyon is the
only propagating degree of freedom.  From the infinite ladder of
string states, only some special discrete states at discrete
euclidean momenta remain.  These lead to short distance bulk
interactions between the tachyon quanta, described at the lowest
order by dilaton-gravity.  Tachyon pulses enter from $\cal{I}$$^-$
to be scattered by the tachyon wall and return to $\cal{I}$$^+$.
In the bulk, these pulses can interact with each other via
either tachyon three-point (and higher) vertices, or by exchanging
gravitons and dilatons (and more massive string fields).
We will reconstruct the gravitational interaction between these
pulses and the resulting metrics from boundary data alone.

The matrix model, and the corresponding collective field theory, 
can be thought of as providing boundary data for tachyon scattering.  
In particular, together with the leg-pole transform, the matrix model allows
us to calculate the exact shape of the outgoing tachyon pulse given
the incoming pulse.  It is in this sense that the matrix model
provides us with a holographic description of dilaton-gravity.
The two dimensional gravity background has two null boundaries, 
$\cal{I}$$^\pm$, and the one dimensional gauge theory supplies 
the scattering matrix between them.

At the same time, we have an equivalence between two different
theories of a scalar field in 1+1 dimensions: one with gravity
(dilaton-gravity interacting with the tachyon field)
and one without (collective field theory for MQM).
We will see in detail how it is possible for these
two theories to be equivalent, shedding perhaps some light on
how gravity can be an emergent theory.

Naively, the spacetime on which the collective field theory
lives is fixed.  As was first pointed out in \cite{Karczmarek:2003pv},
this is not the case.  Sufficiently large fluctuations of the Fermi
sea of eigenvalues can in fact make the spacetime on which
the collective field lives time-dependent, changing its structure, 
for example, by introducing  space-like
infinities \cite{Das:2004aq}.  In such scenarios, it would be most
interesting to be able to calculate the metric and dilaton of the
equivalent string theory target space.  We will see that our
methods make this partially possible, and will calculate the 
metric and the dilaton for a particular time-dependent
scenario.

We work in the convention where $\alpha' = 1$.  In order to avoid
complications arising from the tachyon background, we will be
focusing on bulk processes which do not involve it.  To separate
these bulk processes from the interactions with the background,
we will make the background parametrically small by
taking the string coupling at the tachyon wall to be large.

The reminder of the paper is organized as follows:
In Section \ref{s2} we will discuss exact solutions to the collective
field equations of motion, in the chiral, or lightcone, formalism.
In Section  \ref{s3} we will solve the equations of motion of
dilaton-gravity coupled to a scalar 
perturbatively to third order in the scalar field.
In  Section  \ref{s4} we will tie the results of two previous
sections together, and extract some information about the relationship
between the tachyon and the collective field of matrix eigenvalues. 
Finally, in  Section  \ref{s5}, we will use our techniques to study
a particular time-dependent background.
 
\section{Solutions in chiral formalism}
\label{s2}

In this section, we will obtain explicit formulas linking the 
profile of outgoing fluctuations in the collective field to the profile
of incoming fluctuations.  We will start with a brief review
of the salient facts and definitions in the matrix model. 
For more details, please see 
\cite{Klebanov:1991qa}, \cite{Ginsparg:1993is}, and
\cite{Polchinski:1994mb} (chapter 5).

The $c=1$ MQM has as its fundamental
degrees of freedom non-interacting fermions in the upside down 
harmonic oscillator potential, with the Hamiltonian
\be
H = \half p^2 - \half x^2~.
\label{hamiltonian}
\ee
The curvature of the potential is fixed by taking $\alpha'=1$.
The effective (or bosonized) picture for this system is that of
a Fermi fluid moving in phase space $(x,p)$.  
Due to the incompressibility of
this fluid, in the classical limit 
it is sufficient to give the position of the 
Fermi surface, often presented as $p_\pm(x,t)$, the upper
and lower branches of the Fermi surface as a function of $x$.
The local density of fermions is then given by the distance between the
two branches of $p$:
\be
\varphi(x,t) \equiv {1\over 2} \left ( p_+(x,t) - p_-(x,t)\right)~. 
\ee
The static Fermi surfaces are constant energy hyperbolas given by 
$E = \half p^2 - \half x^2 = \mu$.  For $\mu<0$ the left and the 
right branches of the hyperbola do not interact:
any small fluctuation around this static background evolves
by moving along one arm  from $x=\infty$ towards
finite $x$ and back out to $x=\infty$ along the other arm of the 
same branch.

This description is unfortunately singular, at the place where
the upper and the lower edge of the Fermi sea join.  To avoid this
singularity, we will use an equivalent description with
$\mu > 0$ and allow the fluctuation to propagate from the left 
to the right along the upper branch of the hyperbola $p^2 -  x^2 = 2\mu$.

To calculate the relationship between the incoming and the outgoing
pulse, we will use the chiral (or lightcone) formalism 
\cite{Alexandrov:2002fh}.
The chiral formalism is based on the simple observation that
if the Fermi surface is given at t=0 by $G(x,p)=0$ for
some function $G$, then the time evolution of this Fermi surface is
given by
\be
\label{eqn:general-soln}
G\left ( \half (p+x)e^{-t} - \half (p-x)e^{t} ,
\half (p+x)e^{-t} + \half (p-x)e^{t} \right ) = 0~.
\ee

Consider now the fluctuations of the upper branch of the hyperbola given by
$p^2 - x^2 = 2\mu$, with $\mu>0$. We define the fluctuation field $\eta$ with
\be
p(x,t) = \sqrt{2\mu + x^2} + 2 \sqrt{\pi} ~\partial_x \eta ~.
\label{eqn:def-eta}
\ee
It will turn out that $\eta$ is best thought of as a function of
$\sigma$ such that $x = \sqrt{2 \mu} \sinh \sigma$, so that
$ \partial_x \eta = (2\mu + x^2)^{-1/2}~\partial_\sigma \eta(\sigma)
\approx |1/x|~\partial_\sigma \eta(\sigma)$.
The fluctuations come in from $x \rightarrow -\infty$ and exit at
$x \rightarrow \infty$. Let us see this in detail.  

Consider the following exact time-dependent profile of type 
(\ref{eqn:general-soln}) for some
function $f_{in}(\sigma)<<\mu $, whose meaning will become clear in a moment:
\be
\label{eqn:in}
p^2-x^2 =   2 \mu - 2 f_{in}'
\left ( \ln  \left (  \sqrt{1\over 2 \mu } (p-x)e^t  \right ) \right )~.
\ee
This can be rewritten as
\be
p+x =   {2 \mu \over p-x} -  {2 \over p-x} ~f_{in}'
\left ( \ln  \left (  \sqrt{1\over 2 \mu } (p-x)  \right ) + t \right )~.
\label{eqn:p+x}
\ee
Assuming that $f_{in}(y)$ has  finite support on some interval near $y=0$,
for $t \rightarrow -\infty$, $f_{in}$ in nonzero only if  $p-x$ is 
large.  Then, the right hand side of our equation is small, and
we must have $p \approx -x$.  Therefore, $x$ is large and negative, and 
$p-x \approx -2x$.  Substituting this in, we get
\be
p = -x - {\mu \over x} + {1 \over x} f_{in}'
\left ( \ln  \left (  - \sqrt{2 \over \mu } x  \right ) + t\right )~,
\ee
which for large negative $x$ can be rewritten as
\be
p = \sqrt {x^2 + 2\mu} - { 1\over|x|}~f_{in}'
\left (t-\sigma \right )~,
\ee
with $x =  \sqrt{2 \mu} \sinh \sigma \approx -\sqrt{\mu/2} \exp(-\sigma)$.
We can now identify $f_{in}$ with the early time $\eta$ in equation
(\ref{eqn:def-eta}), which is right-moving as expected.  To be precise,
for $t \rightarrow -\infty$, $\eta(\sigma,t) = (2 \sqrt\pi)^{-1} f_{in}
(t - \sigma)$.

The same analysis applies to late time fluctuations at large positive x.
Starting with a time-dependent profile given by
\be
\label{eqn:out}
p^2-x^2 =   2 \mu - 2 f_{out}'
\left (-\ln  \left (  \sqrt{1\over 2 \mu } (p+x)e^{-t}  \right ) \right )~,
\ee
we can identify  $\eta(\sigma,t) = (2 \sqrt\pi)^{-1} f_{out}(t-\sigma)$
 at late times, $t \rightarrow -\infty$, with 
$x  \approx \sqrt{\mu/2} \exp(\sigma)$.

The crucial observation now is that the profiles in equations
(\ref{eqn:in}) and (\ref{eqn:out}) are exact solutions and valid at all times.
Therefore, if the incoming profile is $f_{in}$, the outgoing profile can
be obtained from setting the right hand sides of equations (\ref{eqn:in}) and 
(\ref{eqn:out}) equal:
\be
 f_{in}'
\left (\ln  \left (  \sqrt{1\over 2 \mu } (p-x)e^{t}  \right ) \right )
=
 f_{out}'
\left (-\ln  \left (  \sqrt{1\over 2 \mu } (p+x)e^{-t}  \right ) \right )~,
\label{eqn:time-delay}
\ee
 We now substitute the expression for x+p
from equation (\ref{eqn:p+x}), and define 
$y = \ln  (   (p-x)e^{t}/\sqrt{2 \mu })$ to get
\be
 f_{in}' (y)= f_{out}' \left (
y - \ln \left ( 1 - \mu^{-1} f_{in}'(y)\right )
\right )~,
\ee
or, defining $y(z)$ by 
$z = y- \ln \left ( 1 - \mu^{-1} f_{in}'(y)\right )$,
$f_{out}' (z)= f_{in}' (y(z)),$
which is nothing more but the time delay equation in 
\cite{Polchinski:1991uq,Natsuume:1994sp}.
More interestingly, we can find the profile at any time.
Given the incoming profile $f_{in}$, equation (\ref{eqn:in})
can be solved for $p$ as a function of $x$ treating
$f_{in}$ as a small perturbation.  $\eta$ as a function of $x$ (or
$\sigma$) and $t$ can then be read off.

To second order in $f_{in}$, we get
\bear
p &=& \sqrt{x^2+2\mu}~-~ {1 \over \sqrt{x^2+2\mu}} f_{in}'(t-\sigma)
\\ \nn &+& {e^\sigma \over \sqrt{2\mu} (2\mu + x^2)}
f_{in}'(t-\sigma) f_{in}''(t-\sigma)\\ \nn &-&
{1 \over 2  (2\mu + x^2)^{3/2}} \left (f_{in}'(t-\sigma) \right )^2
~+~ {\it{o}}((f_{in}')^3)~,
\eear
or
\bear
&& 2\sqrt{\pi}~\partial_\sigma\eta(\sigma,t) = \\
\nn &&
-f_{in}'(t-\sigma) ~+~
{e^\sigma \over 2\mu \cosh \sigma}f_{in}'(t-\sigma) f_{in}''(t-\sigma)
~-~ 
{\left ( f_{in}'(t-\sigma) \right )^2  \over 4\mu \cosh^2 \sigma}
~+~ {\it{o}}((f_{in}')^3)~.
\eear
This can be integrated with respect to $\sigma$
\be
2\sqrt{\pi} ~\eta(\sigma,t) = f_{in}(t-\sigma) ~-~
{e^\sigma \over 4\mu \cosh \sigma} \left ( f_{in}'(t-\sigma) \right )^2 
~+~ {\it{o}}((f_{in}')^3)~.
\label{eta-bulk-2}
\ee

As a consistency check, we notice that in the large negative $\sigma$
regime, $\eta = \eta_{in}$, as expected, and that
in the large positive $\sigma$ regime, there
are no left-moving terms (everything is a function of $t-\sigma$).
In particular, if we take $\sigma \rightarrow \infty$ in the 
above equation, then
\be
2 \sqrt{\pi} ~\eta(\sigma\rightarrow \infty) = f_{in}(t-\sigma) ~-~
{1 \over 2 \mu }\left (f_{in}'(t-\sigma) \right )^2
~+~ {\it{o}}((f_{in}')^3)~.
\ee

To third order, the calculation is a bit more messy.  The result 
for $\eta'$ is again a total derivative, and can be integrated to give
\bear
\nn
2 \sqrt{\pi} ~\eta(\sigma,t) &=& f_{in}(t-\sigma) ~-~
{e^\sigma \over 4\mu \cosh \sigma} \left ( f_{in}'(t-\sigma) \right )^2 
 \\ &-& 
{\left ( e^{2\sigma} + 3\right ) e^\sigma \over 48 \mu^2 \cosh^3(\sigma)}
\left ( f_{in}'(t-\sigma) \right )^3   
 \label{eta}  \\ &+& \nn
 {e^{2\sigma} \over 8 \mu^2 \cosh^2(\sigma)}
\left ( f_{in}'(t-\sigma) \right )^2 f_{in}''(t-\sigma) 
~+~
{\it{o}}((f_{in}')^4)~.
\eear

As a check, let us compute the large $\sigma$ limit of this expression
from the time delay equation, 
$y = z + \ln \left ( 1 - \mu^{-1} f_{in}'(y)\right )$. Expanding, we get that
\be
y(z) = z - \mu^{-1}f_{in}'(z) +  \mu^{-2}f_{in}'(z)  f_{in}''(z) - \half
\mu^{-2}(f_{in}'(z))^2  ~+~ {\it{o}}((f_{in}')^3)~,
\ee
and therefore
\bear
&& f_{out}' (z)= f_{in}' (y(z)) =  
f_{in}'(z) -  \mu^{-1}f_{in}'(z)  f_{in}''(z) + \\ \nn && 
+ \mu^{-2} (f_{in}'(z)  (f_{in}''(z))^2 - \half \mu^{-2} 
 (f_{in}'(z))^2  f_{in}''(z))
+ \half  \mu^{-2} (f_{in}'(z))^2  f_{in}'''(z)
 ~+~ {\it{o}}((f_{in}')^4)~,
\eear
which can be integrated to give
\be
f_{out} = f_{in} - {1\over 2\mu} (f'_{in})^2 
- {1 \over 6\mu^2} (f'_{in})^3 +
{1\over 2\mu^2} f''_{in} (f'_{in})^2 
 ~+~ {\it{o}}((f_{in})^4)~,
\ee
implying that
\be
\eta_{out} = \eta_{in} - {\sqrt{\pi}\over \mu} (\eta'_{in})^2 
- {2\pi \over 3\mu^2} (\eta'_{in})^3 +
{2\pi\over \mu^2} \eta''_{in} (\eta'_{in})^2 
 ~+~ {\it{o}}((\eta_{in})^4)~.
\label{eta-in-out}
\ee

This agrees with equation (\ref{eta}) when $\sigma \rightarrow \infty$.

Our procedure can clearly be extended to any order, and gives both
the collective field profile at any time, and the outgoing profile
for $t\rightarrow \infty$ in terms of the incoming field, as
illustrated in Figure \ref{f1}(b).


\section{Dilaton-Gravity coupled to a massless scalar}
\label{s3}

Having studied the interior behaviour in the collective field
theory, we now turn out attention to the interior behaviour of the
dilaton-gravity theory.

As was described above, effective field theory for Liouville 
string theory is
dilaton-gravity coupled to the (massless) tachyon scalar.
Since the tachyon is a massless field and not actually
tachyonic, the action for these three degrees of freedom is
perfectly well defined.  Denoting the dilaton field with
$\Phi$ and the tachyon with $T$, we have \cite{Natsuume:1994sp}
\be
S = \half \int dt dx 
\sqrt{-G} e^{-2\Phi} \left [ 
a_1 [R + 4(\nabla \Phi)^2 + 16]
-(\nabla T) +4T^2 - 2V(T)
\right ]~,
 \ee
where we will take the tachyon potential to be
\be
V(T) = {a_2 T^3 \over 3}~.
\ee
Here $a_1$ and $a_2$ are constants which were determined in
\cite{Natsuume:1994sp} to be $a_1 = \half$ and $a_2=-2\sqrt 2$.

In conformal gauge, where the metric is $ds^2 = -e^{2\rho} dx^+ dx^-$,
the equations of motion are \cite{Callan:1992rs}
\bear
\label{eq1}
2 \partial_+^2\Phi -  4 \partial_+ \rho \partial_+ \Phi
&=& a_1^{-1}\partial_+ T \partial_+T 
\\
\label{eq2}
 2 \partial_-^2\Phi -
 4 \partial_- \rho \partial_- \Phi
 &=& a_1^{-1}\partial_-T \partial_-T 
\\
\label{eq3}
  2 \partial_+\partial_-\Phi - 4 
\partial_+ \Phi \partial_- \Phi - 4 e^{2\rho} 
&=&  a_1^{-1}e^{2\rho}\left ( T^2 - {a_2 \over 6} T^3 \right ) 
\\
\label{eq4}
4 \partial_+\partial_-\Phi - 4 \partial_+ \Phi \partial_- \Phi 
-2 \partial_+\partial_- \rho - 4 e^{2\rho} &=&
a_1^{-1}   \partial_+T \partial_- T \\ \nn
&+& a_1^{-1}e^{2\rho}\left ( T^2 - {a_2 \over 6} T^3 \right ) 
\\
\label{eq5}
 e^{-2\rho} \left ( \partial_+\partial_-T - \partial_+\Phi \partial_- T 
- \partial_-\Phi \partial_+ T  \right ) - T  &=& -{a_2 \over 4} T^2
\eear
The first three equations are for the metric, the fourth is
for the dilaton and the last is for the tachyon field.
The last two equations can be combined to give a particularly
simple relationship
\be
2 \partial_+ \partial_-(\rho - \Phi) +a_1^{-1}\partial_-T \partial_+T = 0~.
\ee
In the absence of tachyon field, the above equation becomes
$\partial_+ \partial_-(\rho - \Phi)$.  Using up the left-over
coordinate freedom $x_\pm \rightarrow \tilde x_\pm(x_\pm)$,
we could set $\Phi=\rho$, the Kruskal gauge.  However,
since we are dealing with a linear dilaton background, a more
natural gauge would be the modified Kruskal gauge
$ \Phi = x^+ - x^- + \rho$.  Either gauge choice is only
possible in regions where the tachyon field is zero.

We will expand in powers of the tachyon field.  To zeroth order,
we have the linear dilaton background,
\be
\Phi_0 = 2x = x^+ - x^-, ~~~~~~~~~~\rho_0 = 0~.
\ee

The tachyon background is a solution to the linearized version of 
equation (\ref{eq5}) in the linear dilaton background,
\be
  \partial_+\partial_-T -  \partial_- T 
+  \partial_+ T  - T  = 0~.
\label{1st}
\ee
The most general static solution to this equation is
\be
T_0 = (b_1 x + b_2) e^{2x}~.
\ee
We are working in the limit where the tachyon background
can be neglected, $b_1, b_2 \rightarrow 0$, and will be expanding
in powers of the incoming tachyon field: $T =  
T^{(1)} + T^{(2)} + T^{(3)} + ...$, ignoring $T_0$.

It will be convenient to absorb
a factor of the dilaton background into $T$ by defining a new field 
$S = e^{-\Phi_0} T= e^{-2x}T = S^{(1)} + S^{(2)} + S^{(3)} + ...$.
To lowest order the equation of motion is simply
\be
\partial_-\partial_+ S^{(1)} = 0~.
\ee
The rescaled tachyon field $S$ is a massless scalar; above
equation has solutions of the form
$S^{(1)} = S_-^{(1)}(x_-) +  S_+^{(1)}(x_+)$.
Since the region $x\rightarrow +\infty$ is the strong coupling
region, protected by the tachyon condensate, $S$ cannot have
left-moving incoming excitations, and we are left with
$S^{(1)} = S_-^{(1)}(x_-)$.

To second order, we can linearize equations (\ref{eq1}-\ref{eq4})
in gravity and dilaton fluctuations about the background,
$\Phi = \Phi_0 + \delta$, to obtain
\bear
\label{eqq1}
\partial_+^2 \delta - 2\partial_+ \rho
&=& {1 \over 2a_1} (\partial_+T^{(1)})^2 \\
\label{eqq2}
\partial_-^2 \delta + 2\partial_- \rho 
&=& {1 \over 2a_1} (\partial_-T^{(1)})^2 \\
\label{eqq3}
\partial_+ \partial_- \delta - 4 \rho +2 \partial_+ \delta
- 2\partial_- \delta
&=& {1 \over 2a_1} (T^{(1)})^2 \\
\label{eqq4}
2 \partial_+ \partial_- \delta + 2 \partial_+ \delta
- 2 \partial_-\delta - \partial_-\partial_+ \rho - 4\rho 
&=& {1 \over 2a_1} \left ( \partial_+T^{(1)} \partial_-T^{(1)} +
(T^{(1)})^2 \right )~.
\eear
The tachyon equation of motion at this level is
\be
  \partial_+\partial_-T^{(2)} -  \partial_- T^{(2)} 
+  \partial_+ T^{(2)}  - T^{(2)}  = -{a_2 \over 4} (T^{(1)})^2 
\ee
or
\be
\partial_-\partial_+ S^{(2)} = -{a_2 \over 4} e^{2x} (S^{(1)})^2~.
\ee
This last equations is easy to solve for $S^{(2)}(x_-)$
\be
\label{S2}
S^{(2)} = -{a_2 \over 4} ~e^{x^+} ~\int^{x^-} dx^-~ e^{-x^-} (S^{(1)}(x^-))^2~.
\ee
Defining $\Omega = 2(\partial_- - \partial_+)\delta + 4\rho$, we can 
combine equations (\ref{eqq1}-\ref{eqq3}) into
\bear
(\partial_+ - 2) \Omega &=& {1 \over a_1} 
\left ( (T^{(1)})^2 - (\partial_+T^{(1)})^2 \right )\\
(\partial_- + 2) \Omega &=& {1 \over a_1} 
\left ( -(T^{(1)})^2 + (\partial_-T^{(1)})^2 \right )\\
\partial_+ \partial_- \delta 
&=& \Omega + {1 \over 2a_1} (T^{(1)})^2 
\eear
while equations (\ref{eqq3}) and (\ref{eqq4}) give
\be
\partial_+ \partial_- (\delta -\rho) = 
{1 \over 2a_1}\partial_+T^{(1)} \partial_-T^{(1)}~.
\ee
These four equations can be integrated explicitly to give
$\rho$ and $\delta$.  They are more equations (four) 
than unknown functions (two), consistency requires that
$T$ satisfy the 1st order equation (\ref{1st}), whose most
general solution is
\be
T^{(1)} =
\sqrt{a_1}~ e^{x^+ - x^-} \left ( f_+(x^+) +  f_-(x^-)\right )~.
\ee
It is easy to show that in that case, the first two equations
give
\footnote{Note on notation: anytime a limit is not shown for an integral,
it is $+\infty$ for an upper limit, and $-\infty$ for a lower limit.
Integrals with no limits at all should be interpreted as
being over the entire real line.}
\be
\Omega = -e^{2x^+ - 2x^-} \left ( f_+^2 + 2f_-f_+ + f_-^2 -
\int_{x^+} dx^+ ~\left ( f'_+\right )^2 
- \int^{x^-}dx^- ~ \left ( f'_-\right )^2 + 4A
\right )~,
\label{soln:omega}
\ee
and the third gives
\bear
\label{soln:delta}
\delta = 
&~&{1\over 4}  \int^{x^-} dx^- ~e^{2x^+ - 2x^-}
\left [(f'_-)^2 - f_-^2  \right ]
~+~ {1\over 4}  \int_{x^+} dx^+ ~e^{2x^+ - 2x^-} 
\left [(f'_+)^2 - f_+^2  \right ]
 \nn \\ 
&-&{1\over 4} e^{2x^+ - 2x^-} \int_{x^+} dx^+ ~ (f'_+)^2 
~-~ {1\over 4}  e^{2x^+ - 2x^-}\int^{x^-} dx^- ~  (f'_-)^2 
\\ \nn &+& \int^{x_-} dx^-~e^{-2x^-} f_- ~ \int_{x_+} dx^+~e^{2x^+} f_+
- A e^{2x^+ - 2x^-} + \a_+(x^+) - \alpha_-(x^-)
~.
\eear
From the definition of $\Omega$, $\rho$ is then
\bear
\label{soln:rho}
\rho &=& {1\over 4}  \int^{x^-} dx^- ~ e^{2x^+ - 2x^-} 
\left [ (f'_-)^2 -f_-^2 \right ]
+  {1\over 4}  \int_{x^+} dx^+ ~ e^{2x^+ - 2x^-} 
\left [ (f'_+)^2 -f_+^2 \right ]
\nn \\ \nn
&-& {1\over 8} e^{2x^+ - 2x^-} \left ( f_+^2 + 4f_-f_+ + f_-^2 
+2\int_{x^+}dx^+ ~ \left ( f'_+\right )^2 
+2\int^{x^-} dx^- ~\left ( f'_-\right )^2
\right ) 
\\ &-& \nn \half  e^{2x^+} f_+~ \int^{x_-} dx^-~e^{-2x^-} f_- 
~-~ \half  e^{-2x^-} f_-~ \int_{x_+} dx^+~e^{2x^+} f_+
\\  &-& A e^{2x^+ - 2x^-} + \half \partial_+\a_+(x^+) +
\half \partial_- \alpha_-(x^-)~.
\eear
and the fourth equation is satisfied automatically (it is in fact
implied by the other three combined with (\ref{1st})). 

In the above solution, $A$ is an arbitrary integration constant, and
$\alpha_\pm$ are arbitrary integration functions.
$\alpha_\pm$ can be removed from the solution by a coordinate
transformation which respects conformal gauge, namely (to linear order)
$x^\pm \rightarrow x^\pm + \alpha_\pm(x^\pm)$.  In the interest of
simplicity, we will adopt a coordinate system where $\alpha_\pm=0$ for
now, and return to the issue of coordinate ambiguity later.

In contrast with $\alpha_\pm$, the constant $A$ cannot be set
to zero by a coordinate change.  Its presence, however, 
is contrary to our implicit boundary conditions, since $A e^{2x^+ - 2x^-}$
is large for $x^- \rightarrow -\infty$.  If we imagine that the
incoming tachyon pulse is localized (as shown in Figure \ref{f1}(a)),
the metric before the pulse arrives should be flat.   As we will see
in a moment, inclusion of  a nonzero $A$ corresponds to a black hole
background.  We will therefore set $A=0$ as well.  Similar 
arguments apply to the region of spacetime where 
$x^+ \rightarrow \infty$.

The general solution in equations 
(\ref{soln:omega},\ref{soln:delta},\ref{soln:rho}) 
 is more than what we require.
Because our theory has only one asymptotic weakly coupled region,
and because we have ignored the presence of the background which
can reflect back a scalar pulse,
$S^{(1)}$ has only one component, and not two: $S^{(1)} = S^{(1)}(x_-)
= \sqrt{a_1}f_-(x^-)$.  Therefore (dropping the $(1)$ subscript on S for
brevity),
\be
\Omega = -{1\over a_1}~e^{2x^+ - 2x^-} \left ( S_-^2 
- \int^{x^-}dx^- ~ \left ( S'_-\right )^2
\right )~,
\ee
\be
\delta = 
 {1\over 4 a_1}  \int^{x^-} dx^- ~ e^{2x^+ - 2x^-} 
\left [ (S'_-)^2  -S_-^2 \right ]
~-~{1\over 4 a_1} e^{2x^+ - 2x^-} \int^{x^-} dx^- ~ (S'_-)^2 
\label{delta}
\ee
and 
\bear
\label{rho}
\rho &=& {1\over 4 a_1}  \int^{x^-} dx^- ~ e^{2x^+ - 2x^-}
\left [ (S'_-)^2  -S_-^2 \right ] \\ \nn
&-& {1\over 8 a_1} e^{2x^+ - 2x^-} \left (  S_-^2 
+ 2\int^{x^-} dx^- ~\left ( S'_-\right )^2
\right )~.
\eear

If the  incoming pulse is localized around some $x^-$,
and we look at larger values of $x^-$, the metric and the dilaton
outside the pulse simplify to
\be
\delta = \rho =
 {1\over 4 a_1} e^{2x^+} \int dx^- ~ e^{ - 2x^-} 
\left [ (S'_-)^2  -S_-^2 \right ]
~-~{1\over 4 a_1} e^{2x^+ - 2x^-} \int dx^- ~ (S'_-)^2 ~,
\label{delta-rho}
\ee
which imply
\be
\Omega = {1\over a_1}~e^{2x^+ - 2x^-} 
 \int dx^- ~ \left ( S'_-\right )^2~.
\ee

Notice that this is nothing else but the standard 2D black hole, which
in Kruskal gauge $\rho = \Phi$ is given by
\be
e^{-2\rho} = e^{-2\Phi} = {m \over 2} -  4(y^+ - y_0^+)
(y^- - y_0^-)~.
\ee
Changing variables $y^\pm - y_0^\pm = \mp e^{\mp2\tilde y^\pm}$ and 
linearizing, we get
\be
e^{2\rho}  = 1 - {m\over 8}e^{2 \tilde y^+ - 2 \tilde y^-} 
\mbox{~~~and~~~}
e^{2\Phi} = {1\over 4} e^{2 \tilde y^+ - 2 \tilde y^-} 
\left ( 1 -{m\over 8} e^{2 \tilde y^+ - 2 \tilde y^-} \right )
\ee
or
\be
\rho  =  -{m\over 16} e^{2 \tilde y^+ - 2 \tilde y^-} 
\mbox{~~~and~~~}
\Phi = \mathrm{const.} +   (\tilde y^+ -  \tilde y^-)
 -{m\over 16} e^{2 \tilde y^+ - 2 \tilde y^-} ~.
\label{blackhole}
\ee
To compare with equation (\ref{delta-rho}), 
let $\tilde y^+ = x^+ + Ce^{2x^+}$ and
$\tilde y^- = x^-$.  Then, for large negative $x^+$,
 the metric and the dilaton in 
(\ref{delta-rho}) and (\ref{blackhole}) agree,
with 
\be
m = {4\over a_1} \int dx^- ~ (S'_-)^2 \mbox{~~~and~~~}
C = {1\over 4 a_1} \int dx^- ~ e^{ - 2x^-} 
\left [ (S'_-)^2  -S_-^2 \right ]~.
\ee
Notice that the mass is simply the integral over the stress energy
of the incoming pulse, as expected, and that, had we included the
integration constant $A$, it would have contributed to 
the mass, signaling the presence of an undesirable black 
hole background unrelated to the tachyon pulse.

To third order, we only need the tachyon equation,
which now includes interactions with the metric and the 
dilaton,
or
\be
\partial_-\partial_+ S^{(3)} = -{a_2 \over 2} e^{2x} (S^{(1)}) (S^{(2)}) 
+ \half \Omega S^{(1)} 
+ \partial_+ \delta \partial_- S^{(1)}  + \partial_- \delta \partial_+ S^{(1)} 
~.
\ee

With the explicit forms of $\Omega$ and $\delta$
above, this equation can  be integrated as well.  Let us
treat a special case, where we will imagine that the 
incoming field is made up of two well separated pulses
with finite support, $S^{(1)}_-(x^-) =  S^{(1A)}_-(x^-) +  S^{(1B)}_-(x^-)$,
with the A pulse centered around $x^-_A$ and the B pulse 
centered around $x^-_B = x^-_A + T$, with $T$ large.
We will think of $ S^{(1A)}_-(x^-)$ as a source for the
second order fields (tachyon, dilaton and metric) and 
examine scattering of the second pulse, B, from this
background.  For $x_- \rightarrow +\infty$, the
outgoing third order tachyon field is
\bear
S^{(3)} &=&  -{a_2 \over 2} \int^{x^+} dx^+ \int dx^- ~
e^{x^+ - x^-} (S^{(1B)}_-) (S^{(2)A}_-) \\
&+&  \int^{x^+} dx^+ \int dx^- ~
\left (\half \Omega  - \partial_+ \partial_- \delta \right )
S^{(1B)}_-~.
\eear
Combining all our previous results,
\bear
&&-{a_2 \over 2} e^{x^+ - x^-} (S^{(2)A})+
\half \Omega  - \partial_+ \partial_- \delta  = 
\\ && \nn
{a_2^2 \over 8}  e^{2x^+ - x^-} 
~\int dx^-~ e^{-x^-} (S^{(1A)}(x^-))^2
-{1\over 2a_1}~e^{2x^+ - 2x^-} 
 \int dx^- ~ \left ( \partial_{x^-}S^{(1A)}_-\right )^2 ~,
\eear
and therefore
\bear
\label{S3}
S^{(3)} &=& {a_2^2 \over 16} e^{2x^+}
\int dx^- ~ e^{ - x^-} S^{(1B)}(x^-)
\int dx^-~ e^{-x^-} (S^{(1A)}(x^-))^2 \\
&-&{1\over 4a_1}~e^{2x^+} \int dx^- ~ e^{ - 2x^-} S^{(1B)}(x^-)
\int dx^- ~ \left ( \partial_{x^-}S^{(1A)}_-(x_-)\right )^2 ~. \nn
\eear

\begin{figure}
\includegraphics[scale=0.9]{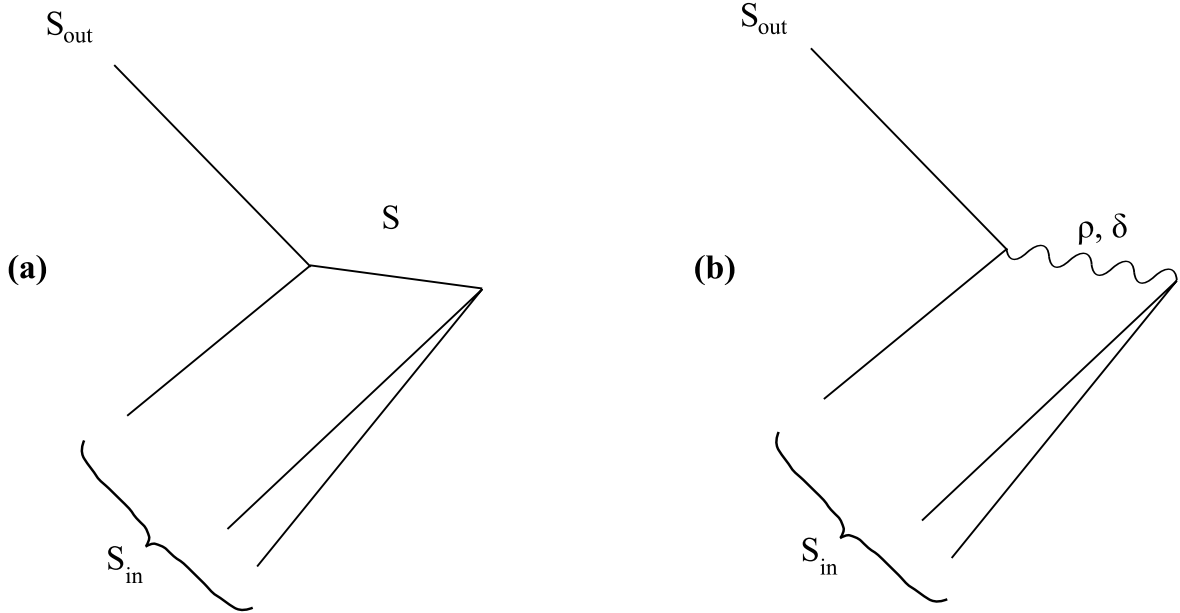}
\caption{Feynman diagram for the third order scattering
of tachyons: {\bf (a)} with a tachyon as an intermediate state
{\bf (b)} with a dilaton or a graviton as an intermediate state.}
\label{f2}
\end{figure}

The first term is due to a Feynman diagram shown in Figure \ref{f2}(a)
and the second due to that in Figure \ref{f2}(b).  In the latter
case, it is the total stress energy of the pulse which determines
the result, in other words, pulse B scatters from the dilaton-gravity
background created by the first pulse.

\section{Relationship between spacetime and matrix model}
\label{s4}

In this section, we will confirm that the results of the 
two preceding sections are linked by the leg-pole transform
on the boundary, and discuss a strategy towards extending
the correspondence into the bulk.

The leg-pole transform  connects the incoming 
tachyon field profile to the incoming collective field
via \cite{Natsuume:1994sp}
\bear
\label{in1}
S_{in}(x^-) &=& -\int dv ~K(v - x^- ) ~\eta_{in}(v) \\
\eta_{in}(\sigma^-) &=& -\int dv ~K(\sigma^- - v ) ~S_{in}(v)
\label{in2}
\eear
and the outgoing fields via
\bear
\label{out1}
S_{out}(x^+) &=& \int dv ~K(x^+  - v  + \ln( \mu/2)) ~\eta_{out}(v) \\
\eta_{out}(\sigma^-) &=& \int dv ~K(v - \sigma^- + \ln( \mu/2) ) ~S_{out}(v)~.
\label{out2}
\eear
$x^{\pm} = t \pm x$ and  $\sigma^\pm = t \pm \sigma$ are
lightcone coordinates in spacetime and in the collective field
theory, as shown in Figure \ref{f1}. The kernel $K$ 
of the leg-pole transform  is given by
\be
K(v) = -{w \over 2} J_1(w)~,~\mathrm{where}~~w = 
2 \left ( 2 \over \pi \right )^{1/8} e^{v/2}
~. \ee
It derives its name from the poles in its Fourier transform,
\be
K(\omega) = \int dv ~e^{-i\omega v}~K(v) =
\left ( 2 \over \pi \right )^{i\omega/4} 
{\Gamma(-i\omega) \over \Gamma(i\omega)}~.
\ee
The frequency space expression was originally derived by
comparing the S-matrix of the matrix model with 
world-sheet results in Liouville string theory 
\cite{Polyakov:1991qx,DiFrancesco:1991ss}.

The Appendix collects some useful formulas about the leg-pole
kernel, which are easily derived in frequency space.

The shifts of $\ln(\mu/2) $ in the outgoing formulas (\ref{out1}) and 
(\ref{out2}) are
related to the position of the tachyon wall.  We will be
taking $\mu \rightarrow 0$, which takes the tachyon wall
 deeply into the strong coupling region and allows us
to neglect, for the most part, scattering from the tachyon 
background.

Minus signs in equations (\ref{in1}) and (\ref{in2}) are 
a result of using the positive energy hyperbola in Section
\ref{s2}.

We combine the scattering formula (\ref{eta-in-out}) 
with the leg-pole transforms to compute the
scattering of the field $S$.  We will do this order by order, and
compare with the results of the previous section.
We will assume that the incoming pulse is well localized 
(with Gaussian fall-off, for example) around $x^- = 0$.

To first order in $S_{in}$ (and $\eta_{in}$) we have
\be
S^{(1)}_{out}(x^+) =  
- \int du ~\left (\int dv  ~K(x^+ + \ln(\mu/2) - v ) K(v - u ) 
\right )~S_{in}(u)~.
\ee
The kernel in the bracket can be, at first approximation, thought to
be local, and centered around $u = x^+ + \ln (\mu/2)$.  Therefore,
for an incoming pulse centered around $x^-$, the bulk of the 
outgoing pulse is centered around $x^+ = x^- - \ln(\mu/2)$, indicating
that the scattering takes place at string coupling
$e^{2x} = e^{x^+-x^-} = 1/\mu = g_{st}$, as expected
(up to a coupling independent shift).
The shape of the tachyon background can be deduced from
the detailed shape of the scattered pulse, but we are not 
interested in it.

To second order, we have
\bear
S^{(2)}_{out}(x^+) &=&  \int dv ~K(x^+ + \ln(\mu/2) - v ) ~
\left (-{\sqrt{\pi}\over \mu} \eta'_{in}(v)^2 \right ) \\ \nn
&=& 
- {1\over 2 \mu} \int du_1 du_2 ~\left (\int dv  ~K(x^+ + \ln(\mu/2) - v ) 
K'(v - u_1 ) K'(v - u_2 ) \right )
\\ \nn && \quad\quad\quad \times S_{in}(u_1) S_{in}(u_2)
\eear
We are interested in the region where $x^+ + \ln\mu$
is large and negative, so we can use formula (\ref{3K}) in the Appendix
to obtain
\be
S^{(2)}_{out}(x^+) =
{\sqrt{2} \over \mu}  \int du~ e^{x^+ + \ln(\mu/2) - u} (S_{in}(u))^2
={1\over\sqrt{2}}   \int du~ e^{x^+  - u} (S_{in}(u))^2~.
\ee
Notice that the answer is independent of $\mu$: this is
bulk scattering, and does not depend on the position of
the tachyon wall.
The answer  is in agreement with equation (\ref{S2}),
with $a_2 = -2\sqrt {2}$. 

To third order, we will assume that the incoming tachyon
profile is made up of two pulses, just like we did in Section 2.  Then,
\bear
S^{(3)}_{out}(x^+) &=& \int dv ~K(x^+ + \ln(\mu/2) - v ) ~
\left ({2\pi\over 3 \mu^2} (\partial_v-1) (\eta'_{in}(v))^3 \right ) \\ \nn
&=& 
{2\pi\over 3 \mu^2} \int du_1 du_2 du_3 ~
\left ( 3~S^B_{in}(u_1) ~S^A_{in}(u_2) ~S^A_{in}(u_3)\right ) 
\\ &&\times
\left (\int dv  ~(1-\partial)
K(x^+ + \ln(\mu/2) - v ) 
K'(v - u_1 ) K'(v - u_2 ) K'(v - u_3 ) \right )
~. \nn
\eear
Using equation (\ref{4K}) in the Appendix, this becomes,
\bear
&&S^{(3)}_{out}(x^+) =  \\ \nn
&& \half \int du_1 du_2 du_3 ~ 
\left (e^{2x^+ - u_1 -u_2} \delta(u_2-u_3) + e^{2x^+ - 2u_1}
\delta''(u_2-u_3) \right)  S^B_{in}(u_1) S^A_{in}(u_2) S^A_{in}(u_3)=
\nn \\ \nn &&
\half  \int du_1 ~e^{2x^+ - u_1} S^B_{in}(u_1)
\int du_2 e^{-u_2}  (S^A_{in}(u_2))^2 
-\half \int du_1 ~e^{2x^+ - 2u_1} S^B_{in}(u_1)
\int du_2    (\partial S^A_{in}(u_2))^2
\eear
which agrees with equation (\ref{S3})
for $a_1=\half$ and $a_2 = -2\sqrt {2}$, as before.

Now that we know the boundary profiles match up, we can
explore the relationship between bulk fields.
From equation (\ref{S2}), using our established value for
$a_2$, we have, to second order
\be
S(x^-,x^+) = S_{in}(x^-) ~+~
 {1\over\sqrt 2} ~e^{x^+} ~\int^{x^-} dx^-~ e^{-x^-} (S_{in}(x^-))^2~.
\label{Sin-S}
\ee
We can try to obtain the relationship between $S$ and $\eta$
at the same time $t$ up to second order by solving for $S_{in}$
in terms of $\eta(\sigma_+, \sigma_-)$ at time $t$.
Concretely, equation (\ref{eta-bulk-2}) implies that, to
second order in $\eta$,
\be
\eta_{in}(\sigma-t) = \eta(\sigma,t) ~+~
{e^\sigma \over 4\mu \cosh \sigma} \left ( \partial_{\sigma^-}
\eta(\sigma,t) \right )^2 
~+~ {\it{o}}(\eta)^3)~.
\ee
While on the surface it would appear that the right hand side
depends on both $\sigma^-$ and $\sigma^-$, we know that this 
is not the case.
We will now apply the equal-time leg-pole transform to obtain
$S_{in}$
\bear
S_{in}(t-x) &=& \int dv ~K(-v + x ) ~\eta_{in}(\sigma^- = t-v)
\\ \nn &=&
\int dv ~K(-v + x )~ \left ( 
\eta(v,t) ~+~ {e^v \over 4\mu \cosh v} \left ( \partial_{\sigma^-}
\eta(v,t) \right )^2 \right )
\eear
which we can then plug into equation (\ref{Sin-S}).

This procedure can be extended to higher orders in
perturbation theory, and would allow us to relate the metric
and the dilaton, as well higher order corrections to the 
tachyon field to the collective field, via equations
(\ref{rho},\ref{delta},\ref{S3}).  Therefore, at least in
principle, we can write the explicit field redefinition linking 
dilaton-gravity coupled to a massless scalar to a theory
with only a single scalar field.
Notice though that going beyond the third order would require
the inclusion of effects of heavy string states into the gravity action.
The map is nonlocal, which should come as no surprise, since 
it can be interpreted as a result of integrating out
dilaton-gravity.

Due to diffeomorphism invariance in the dilaton-gravity theory,
this field redefinition cannot be unique.  We have fixed the
coordinate invariance by asking the fields to be related at equal
times, hence picking a particular coordinate system in the gravitating
theory.  However, for localized
pulses, the ambiguity results in at most exponentially small corrections
at large $x$.

The argument for this last fact rests on form of the detailed agreement
between the collective field and the tachyon on the boundary.
Let us assume a well localized incoming tachyon  pulse.  Under the transform
(\ref{in2}), for $\sigma^-$  large and negative,
 the incoming collective field has a form
$\eta_{in} =  A_1 e^{\sigma^-}  +A_2 e^{2\sigma^-} + \ldots$, 
while for $\sigma^-$  large and positive, the fall-off is much more rapid.
The outgoing collective field has the same form.  Therefore,
the outgoing tachyon field for large negative $x^+$ must also be sum of
terms of the form $e^{kx^-}$ with $k$ a positive integer.
Now, consider the effect of a change of coordinates $x^+ \rightarrow
\tilde x^+$ on equations (\ref{S2}) and (\ref{S3}).  For these
equations to only contribute terms in the form  $e^{k\sigma^-}$,
the change of coordinates much be limited to $x^+ \rightarrow x^+
+ \sum_k B_k e^{kx^-}$.  Therefore, on the boundary the coordinates
can be fixed up to exponentially small ambiguity.  A similar argument
holds for the incoming boundary, and the coordinate $x^-$.

In the bulk, the coordinate changes are limited to those which maintain
conformal gauge.  This is because at the lowest order, both the collective
field and the rescaled tachyon field $S$ are massless scalars,
with an equation of motion $\partial_+ \partial_- S =  
\partial_+ \partial_- \eta = 0$.
To maintain conformal gauge, the bulk coordinate changes must be
of the form $x^\pm \rightarrow X^\pm(x^\pm)$, where the functions
$X^\pm$ must be of the form discussed in the previous paragraph,
and are fixed up to exponentially small corrections.

While explicit, the procedure for connecting bulk fields in the two theories
described in this section is not straightforward.  In the next
section, we will simply use our results from Section \ref{s3} to
discuss some interesting examples beyond localized pulses.

\section{Example: time-dependent background}
\label{s5}

In this section we will employ our results to make a connection
between the matrix model and spacetime physics in a time-dependent
scenario.  For convenience, especially when comparing our results
with previous work on this background 
\cite{Karczmarek:2003pv,Karczmarek:2004ph,Das:2004hw,Mukhopadhyay:2004ff,Ernebjerg:2004ut},  
in this section  we take our matrix
model background to be the left branch of $x^2-p^2 = 2\mu$,
and define the fluctuation field $\eta$ in the standard way \cite{Das:1990kaa},
$(p_+ - p_-)/2 = \sqrt{x^2-2\mu} + \sqrt{\pi}\partial_x \eta$,
which is compatible with our definition in equation (\ref{eqn:def-eta}).
On the left branch of the hyperbola, we have $x = -\sqrt{2\mu} \cosh \sigma$
and we will take $\sigma$ to be negative, so that for large
$x$, $x \approx -\sqrt{\mu/2} \exp(- \sigma)$.

We will focus on the following exact time-dependent 
profile in eigenvalue phase space
\be
(x+p + \lambda e^t)(x-p) = 2\mu~,
\ee
which at large $x$ and large negative $t$ corresponds to 
$\eta \approx -{\lambda \over 2\sqrt\pi} e^t x \approx
 {\lambda \over 2}\sqrt{\mu/2\pi} e^{t- \sigma}$.
This is the incoming $\eta$ profile.

As has been shown in 
\cite{Karczmarek:2004ph,Ernebjerg:2004ut}, the exact effective action for the 
fluctuation $\eta$ in the background with $\lambda\neq 0$ 
is the same as the effective action in the static background
($\lambda =  0$) under a change of coordinates from $\sigma$
to $\tilde \sigma$ given by $\sqrt{2\mu}\cosh\sigma =
\sqrt{2\mu}\cosh\tilde\sigma + (\lambda/2) e^t$.  For $\sigma$ and
$\tilde \sigma$ large and negative, the change of coordinates is
\be
e^{-\sigma^+} = e^{-\tilde \sigma^+} + \tilde \lambda
\label{change-}
\ee
with $\tilde \lambda = \lambda / \sqrt{2\mu}$, $\sigma^+ = t+\sigma$
and $\tilde \sigma^+ = t+\tilde \sigma$.

\begin{figure}
\includegraphics[scale=0.8]{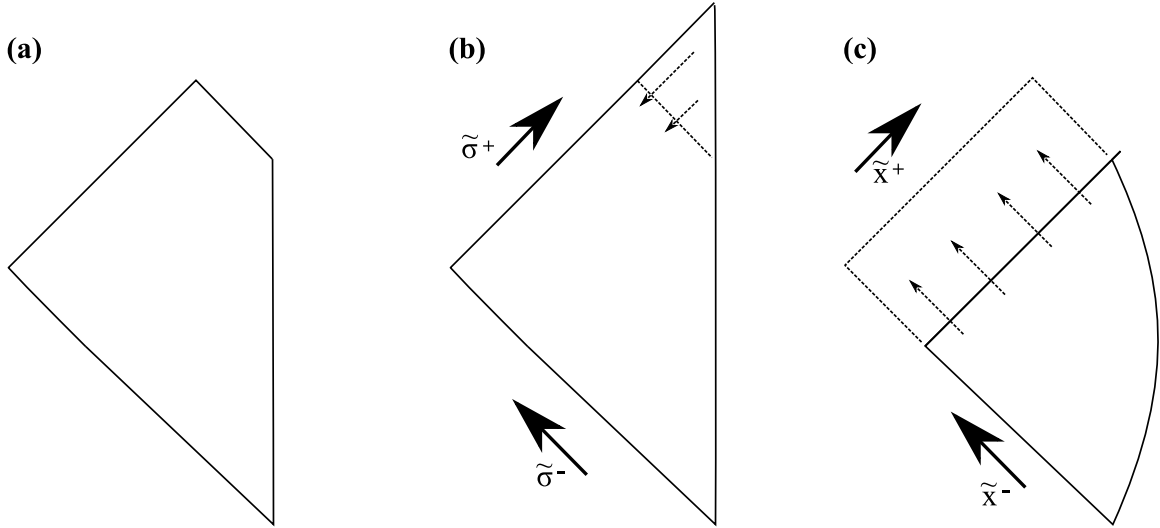}
\caption{Spacetime of the time-dependent solution in different
coordinate systems. {\bf (a)} $\sigma^\pm$ or $x^\pm$ {\bf (b})
 $\tilde \sigma^\pm$
{\bf (c)} $\tilde x^\pm$. Dashed arrows indicate coordinate change relating
the regions in (b) and (c) to the region in (a).}
\label{f3}
\end{figure}

To analyze the spacetime behaviour in this background, we first
notice that under the leg-pole transform, the incoming 
profile $\eta_{in} \sim e^{t-\sigma}$ changes only by an infinite
normalization constant.  We have, therefore,
an incoming field given by $S_{in} \sim e^{x^-}$.  Consider, therefore,
a tachyon background $T_\lambda = \tilde \mu ( e^{x^+ - x^-} + 
\hat \lambda e^{x^+})$
where $\hat \lambda$ is a renormalized constant, and 
we have added back the standard stationary background term to
regularize our problem.  In this background, the form of small fluctuations
must satisfy a linear version of equation (\ref{eq5}),
\be
\partial_+ \partial_- S = -{a_2\over 2} T_\lambda S
\ee
This equation can be transformed into one where the background is simply
$T_0 = \tilde \mu  e^{x^+ - x^-}$ by the following change of variables
\be
e^{-\tilde x^-} = e^{- x^-} - \hat \lambda  x^-
\label{change+}
\ee

Since in the matrix model the time-dependent background is
equivalent to one which is static, we would expect the same to 
be true in dilaton-gravity.  The combined change of coordinates 
(\ref{change-})  and (\ref{change+}) relate these static 
backgrounds to each other, at least at large $x$ (or $\sigma$).
The behaviour near the potential barrier is more complicated,
and hard to study in the spacetime picture since the exact form
of the tachyon potential is not well defined.

The physical picture is illustrated in Figure \ref{f3}.   Figure \ref{f3}(a)
shows the time dependent spacetime generated by the decaying
Fermi sea \cite{Karczmarek:2004ph} whose ${\cal I}^+$ is incomplete.
Figure \ref{f3}(b) shows how this incomplete spacetime is related to
the static spacetime obtained from the collective theory in
the new coordinates.  Figure \ref{f3}(c) shows this relationship for
the dilaton-gravity theory.  The metric and the dilaton are 
trivial in the $\tilde x^\pm$ coordinates to this order.

To next order, we can calculate the second order tachyon field, as well as
the  metric and the dilaton:
\bear
T^{(2)} &=& -{a_2 \over 4} \tilde \mu \hat \lambda e^{2x^+}
\\
\delta &=& -{1 \over 8 a_1} \tilde \mu \hat \lambda e^{2x^+}
\\
\rho &=& -{1 \over 4 a_1}\tilde \mu \hat \lambda e^{2x^+}
\eear
Since the effect of the second order field is 
small in the region where the the coordinate
changes (\ref{change-})  and (\ref{change+}) are nontrivial, they will
not have a large effect on the spacetime analysis we have 
presented already.

\section{Conclusion and further directions}

The results of Section \ref{s3} can be used to rewrite the 
theory of a scalar coupled to dilaton-gravity without the 
need for the dilaton and gravity fields, at least to lowest
order in those fields.  Simply take the expressions 
(\ref{soln:delta}) and (\ref{soln:rho}) and substitute them
back to the tachyon equation of motion, (\ref{eq5}).
The resulting equation of motion is of course nonlocal, 
as is expected when trying to integrate out gravitational
interaction.  Diffeomorphism invariance of the original
theory manifest itself in the presence of the integration
functions $\alpha_\pm(x^\pm)$.

In Section \ref{s4} we outlined a procedure
for relating the solution of this nonlocal action to 
the simpler solutions of the collective field theory,
using the known boundary correspondence.  This is a toy model
for the much more complicated problem of reconstructing the
spacetime dynamics in AdS/CFT.  Our simple
example in Section \ref{s5} demonstrates how our
results can be used to study time-dependent scenarios
in Liouville string theory.  It would be very interesting
to see how these results can be used in more complicated
scenarios, such as those involving space-like future
boundaries \cite{Das:2004aq,Karczmarek:2007ag}.

\section*{Acknowledgments}
This work was completed with support from the Natural Sciences
and Engineering Council of Canada.

\section*{Appendix: Integrals involving the leg-pole kernel K}

Using the Fourier transform form of $K$, it is easy to
show that the following integrals are true.

\be
\int dy~   K(y-x_1) K(y-x_2) = \delta(x_1-x_2)~,
\ee
for any $x_1$ and $x_2$;

\be
\int dy~   K(x-y) K(y-x_1) = \sqrt{2 \over \pi} \left ( 
(x-x_1) ~+~4\gamma - 2 + \ln\sqrt{2\over\pi} 
\right ) e^{x-x_1}~,
\ee
for $x - x_1$ large and negative;

\be
\label{3K}
\int dy~  K(x-y) \partial K(y-x_1) \partial K(y-x_2) =
- \sqrt{2 \over \pi} ~e^{ x - {x_1}} ~\delta(x_1-x_2) 
\ee
for $x - x_i$, $i=1,2$ large and negative;
and finally, for $x - x_i$ large and negative, and with
$x_1-x_2 \gg |x_2-x_3|$,
\bear
\label{4K}
&& \int dy~  (1-\partial)K(x-y) 
\partial K(y-x_1) \partial K(y-x_2) \partial K(y-x_3) =
\\ \nn
&=& {1\over \pi}~ e^{2x - x_1 - x_2} ~\delta(x_2-x_3)
~+~{1\over \pi} ~e^{2x - 2x_1}~ \delta''(x_2-x_3)~.
\eear

\bibliographystyle{JHEP}
\bibliography{my}

\end{document}